# Effect of ion-implantation-induced defects and Mg dopants on thermoelectric properties of ScN

Nina Tureson[1], Marc Marteau[2], Thierry Cabioch[2], Ngo Van Nong[3], Jens Jensen[1], Jun Lu[1], Grzegorz Greczynski[1], Daniele Fournier[4], Niraj Singh,[5] Ajay Soni,[5] Laurent Belliard[4], Per Eklund[1*], Arnaud le Febvrier[1*]

**Affiliation**

[1] Department of Physics, Chemistry and Biology (IFM), Linköping University, SE-581 83 Linköping, Sweden

[2] Institut PPRIME, CNRS-Université de Poitiers-ENSMA, UPR 3346, SP2MI, F-86962 Futuroscope Chasseneuil Cedex, France

[3] Department of Energy Conversion and Storage, Technical University of Denmark, Risø Campus, Fredriksborgsvej 399, Building 779, 4000 Roskilde, Denmark

[4] Sorbonne Université, CNRS, Institut des NanoSciences de Paris, UMR 7588, Paris 75005, France

[5] School of Basic Sciences, Indian Institute of Technology Mandi, Mandi, Himachal Pradesh 175005, India

*Corresponding authors: arnaud.le.febvrier@liu.se, per.eklund@liu.se

**Abstract**

For applications in energy harvesting, environmentally friendly cooling, and as power sources in remote or portable applications, it is desired to enhance the efficiency of thermoelectric materials. One strategy consists of reducing the thermal conductivity while increasing or retaining the thermoelectric power factor. An approach to achieve this is doping to enhance the Seebeck coefficient and electrical conductivity, while simultaneously introducing defects in the materials to increase phonon scattering. Here, we use Mg ion implantation to induce defects in epitaxial ScN (111) films. The films were implanted with $Mg^+$ ions with different concentration profiles along the thickness of the film, incorporating 0.35 to 2.2 at.% of Mg in ScN. Implantation at high temperature (600 ˚C), with few defects due to the temperature, does not substantially affect the thermal conductivity compared to a reference ScN. Samples implanted at room temperature, in contrast, exhibited a reduction of the thermal conductivity by a factor of three. The sample doped with 2.2 at.% Mg also showed an increased power factor after implantation. This study thus shows the effect of ion-induced defects on thermal conductivity of ScN films. High-temperature implantation allows the defects to be annealed out during implantation, while the defects are retained for room-temperature implanted samples, allowing for a drastic reduction in thermal conductivity.

# 1 Introduction

Thermoelectric materials and devices are applied for energy harvesting, converting waste heat (temperature gradients) into useful electricity, as power sources in remote or portable applications using Seebeck effect, and for environmentally friendly cooling using the Peltier effect [1]. The efficiency of a thermoelectric material is connected to the dimensionless thermoelectric figure of merit ($ZT=S^2\sigma T/\kappa$), which consists of the Seebeck coefficient (S), the electrical conductivity ($\sigma$), the thermal conductivity ($\kappa$), and the absolute temperature (T). To enhance the figure of merit of a material, and thus the efficiency, strategic optimizations are required since all parameters (S, $\sigma$ and $\kappa$) are highly interrelated [2,3].

Different approaches are used for improving ZT, including strategies to increase the power factor and reducing the thermal conductivity. Maximizing the power factor includes the search of new materials or optimization of existing ones using approaches such as doping, alloying and/or nanoscale effects (e.g., quantum confinement) [4]. Minimizing the thermal conductivity can be achieved by alloying, forming composites, use of naturally poor thermal conductors such as some layered materials, with soft phonon modes, and nanostructuring of the materials [2,3,5-10]. For nanoscale materials, quantum size effects can affect the density of state at the Fermi level ($E_F$) and thus increase the power factor [5,11]. From bulk (3D) to thin film (2D), the thermal conductivity can be reduced by boundary scattering without reducing the electrical conductivity or power factor. With thin films, similar approaches are used as for bulk materials, and further approaches include superlattices and multilayers [12-15].

In the present study, we investigate an approach to enhance the power factor of thin films by doping in combination with reduction of the thermal conductivity by the creation of defects (point defects and nanoscale line defects). For doping, we use ion implantation instead of directly introducing the dopants while depositing the material. Ion implantation is a commonly used technique for doping of silicon in the semiconductor industry and is suitable for doping thin films or the surface of bulk materials. It is a method known for precise dose control and good reproducibility as well as full range possible implanted elements. Depending on the energy and mass of the implanted ions, different degrees of damage will be induced in the implanted material. Typical collision cascade effects will so create point defects (vacancies, interstitials, vacancy-interstitial pairs, and antisites) and possibly extended defects (dislocations, vacancy clusters, …) which can progressively evolve or disappear during annealing. These evolutions are strongly materials dependent [16]. In contrast to semiconductor industry, these defects and imperfections can be an advantage for thermoelectric materials and increase phonon scattering, leading to a reduced thermal conductivity.

As a possible model system to demonstrate this general idea of reducing the thermal conductivity as well as doping by ion implantation, we choose ScN. Several of the semiconducting transition metal nitrides, in particular ScN- and CrN-based materials have recently emerged as promising thermoelectric materials [13,17-22]. ScN has favorable properties such as high carrier mobility (10-180 $cm^2$ $V^{-1}$ $s^{-1}$), carrier concentration in the range $10^{18}$–$10^{22}$ $cm^{-3}$ [23], low electrical resistivity (~300 $\mu\Omega.cm$) [19] and a narrow indirect band gap of around 0.9 eV [23,24]. In comparison with established thermoelectric materials like PbTe and $Bi_2Te_3$ [2], the power factor of ScN (2.5–3.3 $Wm^{-1}K^{-2}$) [19,25] is in the same order of magnitude [19,25-29]. However, the thermal conductivity is relatively high (10-12 W $m^{-1}$ $K^{-1}$) [25,30] and needs to be minimized for thermoelectric application. Previous studies have shown different approaches for reducing the thermal conductivity of ScN [27,30-32]. For example, the thermal conductivity of ScN was reduced by a factor of five using Nb alloying, but the power factor was degraded, leading to an overall thermoelectric figure of merit similar to that of ScN [27].

The nature of dopants used in the ScN system has to be chosen wisely. More than creating defects in the ScN matrix, the dopant may play an important role on the electronic and/or optical properties of ScN. Kerdsongpanya *et al.* theoretically demonstrated the influence of introduction of impurities in ScN on either N or Sc sites on the density of states around $E_F$ [33]. The desired effect for maximizing the Seebeck coefficient of thermoelectric materials is to have a steep slope of the transport distribution function close to $E_F$. This can be achieved with the presence of impurities or vacancies in the ScN matrix which creates a peak close to $E_F$ [2,34]. Kerdsongpanya *et. al.* proposed magnesium doping in ScN to achieve a peak shift towards $E_F$ [33]. According to first-principles calculations a few percent of Mg doping is enough to induce these effects [33]. Mg contents above 3 at.% shift $E_F$ into the valence band, rendering the material *p* type, as experimentally demonstrated by Saha *et al.* [28,35]. Furthermore, contaminants such as oxygen and fluorine can act as donors in ScN, also leading to a shift of $E_F$ [25,33].

In the present work, epitaxial ScN thin films were grown using DC reactive magnetron sputtering and then implanted with $Mg^+$ ions. Different implantation conditions were tested in order to analyze the influence on the thermoelectric properties of the concentration of dopants but also of the defects created by implantation. A series of samples implanted at room temperature with different doses from 0 to 2.2 at.%, was used for a complete study with the evolution of the thermoelectric properties with the concentration of dopants implanted with defects. The samples implanted at room temperature with an average concentration of 2.2 at.% of Mg exhibited a large decrease in thermal conductivity by 70 % and an increased absolute value of the Seebeck coefficient by 60 %. One of the samples was implanted at high temperature with 2.2 at.% Mg in order to isolate the effect of magnesium doping since a large fraction of the implantation-induced defects might be annealed during the implantation process at high temperature. The same implantation (2.2 at. % Mg) performed at room temperature on a ScN film was used for comparison and for evaluating the effect of irradiation-induced defects on the thermoelectric properties.

## 2 Experimental details

ScN thin films were deposited using DC reactive magnetron sputtering in an ultra-high vacuum chamber (base pressure $10^{-6}$ Pa) with Sc (50-mm diameter, MaTek: Sc 99.5%). The sputtering targets were operated with 125 W under a pressure of 0.27 Pa (2 mTorr) in an Ar/N$_2$ (flow ratio 75% Ar / 25% N$_2$) sputtering-gas mixture. 10 mm x 10 mm one side-polished substrates of Al$_2$O$_3$ (c-cut) (Alineason Materials & Technology) were used. The sapphire substrates were kept at a temperature of 800 °C and under constant rotation during the deposition. Prior to deposition, the substrates were cleaned first for 10 min in acetone in an ultrasonic bath, then in ethanol, and blown dry with a N$_2$-gun. One sample was selected as a reference (labelled REF), the other six samples were implanted with Mg$^+$ ions. The SRIM 2013 software [36] was used to simulate and determine the appropriate ion energies and respective doses needed to obtain the desired Mg concentration in the ScN film (density of 4.26 g/cm$^3$ (calculated from the ICDD data 032-0656286)). Two different profiles for the Mg concentration were applied: one flat by using five different energies for Mg$^+$ ions and one with a Gaussian-like profile along the thickness of the film by using only one energy for implanted ions. The implanted dose of Mg was adjusted for each energy by controlling the duration of the implantation while maintaining a current beam density not exceeding 5μA.cm$^{-2}$ to avoid a temperature increase of the ScN films during the implantation process. Table 1 summarizes the conditions of implantation for the different samples. The first series of samples were implanted at room temperature with different average concentrations of Mg (Mg/(Sc+N+Mg)) from 0.35 at.% to a maximum of 2.2 at.% (labelled RT as Room Temperature). Another film was also implanted to obtain 2.2 at.% of Mg at a temperature of 600 °C (labeled HT as High Temperature). Finally, a sample was implanted at room temperature but using only one energy of 150 keV for implanted Mg$^+$ (labelled SE as Single-Energy) with a total dose selected to obtain an average of 2.2 at.% of Mg in ScN, but with a Gaussian-like profile. For every implantation, the samples were tilted with an angle of 2-5° to prevent channeling of the implanted Mg ions into the epitaxial ScN thin films.

**Table 1.** List of the samples with their labels and different conditions of Mg implantation. The thickness of each film is also listed. The average concentration of magnesium is the one deduced from depth XPS profile measurement performed on the sample 2.2RT and extrapolated to the other samples according the ToF SIMS measurement. (HT = High Temperature, SE = Single-Energy, RT = Room Temperature.)

| Sample label | Average conc. of Mg (at.%) | Thickness (nm) | Temperature (°C) | Fluence of implanted $Mg^+$ ions for the different energies ($\times 10^{15}$ ions/cm$^2$) | | | | | |
|---|---|---|---|---|---|---|---|---|---|
| | | | | 20 keV | 50 keV | 100 keV | 150 keV | 180 keV | Total |
| 2.2 HT | 2.2 | 410 | 600 | 2.3 | 5.6 | 9 | 6 | 20 | 42.9 |
| 2.2 SE | 2.2 | 330 | room temperature | | | | 43 | | 43 |
| 2.2 RT | 2.2 | 345 | room temperature | 2.3 | 5.6 | 9 | 6 | 20 | 42.9 |
| 1.1 RT | 1.1 | 405 | | 1.15 | 2.75 | 4.5 | 3 | 10 | 21.4 |
| 0.75 RT | 0.75 | 410 | | 0.76 | 1.8 | 3 | 2 | 6.7 | 14.3 |
| 0.39 RT | 0.35 | 435 | | 0.38 | 0.9 | 1.5 | 1 | 3.3 | 7.6 |
| REF | 0 | 365 | | | | | | | |

X-ray diffraction (XRD) measurements were performed with an X'Pert PRO from PANalytical apparatus for θ-2θ scans using a Cu K$_\alpha$ radiation with a nickel filter. Philips X'Pert-MRD with Cu K$_\alpha$ radiation was used for the rocking curves and ɸ-scans. Surface and cross sections of the films were examined by scanning electron microscopy (SEM, LEO Gemini 1550, Zeiss). High Resolution Scanning Transmission Electron Microscopy (HRSTEM) images were acquired with the Linköping double C$_s$-corrected FEI Titan$^3$ 60−300 operated at 300 kV using the High Angle Annular Dark Field (HAADF) detector. Time-of-flight Secondary Ion Mass Spectrometry (TOF-SIMS) using a TOF-SIMS V instrument (ION-TOF GmbH, Germany) was used for measuring the Mg distribution in the implanted films. Dual-beam depth profiling, by alternately applying an analysis beam and a sputter beam (non-interlace), was done in positive mode. This allows selected positive secondary ion species to be monitored as a function of sputter time. A low-energy electron flood gun was applied for charge-compensation during profiling. A quasi-continuous 2.0 keV $O_2^+$ beam with a current of 670 nA and scanned over 350 x 350 μm², was used as sputter beam. A pulsed 30 keV $Bi^+$ beam, cycle time 40 μs, was used as analysis beam, with a target current of 3.8 pA and an analysis field of view of 80 x 80 μm² at the centre of the sputter craters. Ion mass spectra were acquired, with an extraction voltage of 2000 V between the sputter sequences, in the so-called spectroscopy mode (bunched 6.5 ns $Bi^+$ ion beam pulse width). SurfaceLab 6 software (v. 6.5, ION-TOF GmbH) was used for spectra recording and data processing. XPS was performed with an Axis Ultra DLD instrument from Kratos Analytical (UK). The system base pressure during spectra acquisition was $1.1\times10^{-9}$ Torr ($1.5\times10^{-7}$ Pa). A monochromatic Al Kα radiation (hν = 1486.6 eV) from the source powered to 150 W was used. Compositional depth profiles were obtained by recording core level spectra after each sputtering step consisting of 3 minutes-long bombardment with 4 keV Ar+ ions followed by 10 min. irradiation at the reduced energy of 0.5 keV to minimize the surface damage and avoid forward implantation of surface species [37]. The Ar+ ion beam was incident at the 20° angle from the surface and rastered over the area of 3×3 mm². All spectra were collected from the area of 0.3×0.7 mm² and at normal emission angle. The analyzer pass energy was set to 20 eV which results in the full width at half maximum of 0.55 eV for the Ag $3d_{5/2}$ peak. Elemental compositions were determined based on Sc 2p, N 1s, O 1s, and Mg 2s peak areas using Casa XPS software (version 2.3.16), and elemental sensitivity factors supplied by Kratos Analytical Ltd.

Thermal conductivity of the films was obtained at room temperature using modulated thermoreflectance microscopy (MTRM). In this setup, a pump beam at 532 nm delivered by a Cobolt MLD laser, intensity modulated by an acousto-optical modulator at a frequency f, is focused on the surface of the sample with an objective lens (N.A. = 0.5). Then, thermal waves were excited in the sample and monitored by the reflectivity surface change recorded around the pump location by another focused laser beam. The specification of the setup is the spatial measurement around the pump beam. We use a 488 nm Oxxius laser to maximize the probe sensitivity to the thermal field on a gold surface. A photodiode and a lock-in amplifier record the AC reflectivity component, in a frequency range between 1 kHz and 1 MHz. The measurement of the reflectivity of the probe on the surface is performed along a x axe from -10 µm to + 10 µm around the pump beam area. The figure S1 represents typical curve of the amplitude and the phase part of the reflectivity signal measured on a gold/substrate and on a gold/film/substrate sample. Finally, the amplitude and phase experimental data were fitted according to a standard Fourier diffusion law to extract the thermal conductivity of the ScN films [38-41]. A full explanation of the thermal conductivity measurement, fitting and model used is reported in the Supplemental Material.

The in-plane Seebeck coefficient and the electrical resistivity were measured simultaneously from room temperature to 500°C under a low-pressure helium atmosphere (~ $9\times10^4$ Pa, purity 99.999% with <0.5 ppm residual oxygen) using ULVAC-RIKO ZEM3 with a special design for thin films. The substrate contribution to the Seebeck coefficient and electrical resistivity is negligible, and the instrumental error is within 7%. The room-temperature Hall effect measurements up to 5 T magnetic field were performed employing physical property measurement system (PPMS Dynacool).

## 3 Results and discussion

XRD θ-2θ scans of as-deposited and Mg-implanted ScN thin films are presented in figure 1a. The observation of only one diffraction peak from the film demonstrated strong (111) texture of the ScN thin films. The inset shows the ScN (111) peaks in magnified view around 34.34°. From here, it is evident that no peak shift is observed after the Mg implantation. The corresponding lattice parameter 4.52 Å is close to earlier reported values (4.50 Å, ICDD PDF 00-045-0978 (ScN)). Figure 1b shows the evolution of the Full Width at Half Maximum (FWHM) of the Rocking curve performed on the 111 reflection of the film. FWHM Values vary between 2.4 ° and 1.9° showing that no degradation of the macroscopic view of the crystal quality of the film was noticeable by XRD. The inset shows a φ–scan of ScN reference sample (χ=70.5°, ScN (111)). The six peaks appear due to twin domain symmetry of ScN grown on sapphire substrates [19,27]. Thus, the films are composed of single phase epitaxial cubic ScN with an out-of-plane [111] orientation. Due to the low quantity of implantation and the small difference of the ionic radius between $Sc^{3+}$(VI) and $Mg^{2+}$(VI), it is not possible to discuss a potential substitution of Mg for Sc by these XRD results.

In figure 2, the optical image and the surface morphology from the SEM of the films are shown. No noticeable change of the morphology of the surface of the film has been observed by SEM after implantation of Mg ions The non-implanted ScN sample has a yellowish color characteristic of ScN material [42], but the Mg-implanted samples are brown/black. This drastic change of color indicates changes in the bandgap with insertion of states or doping [33].

A depth-profile composition analysis of the sample 2.2 RT by x-ray photoelectron spectroscopy gave an average concentration (Mg/(Mg+Sc+N)) around 2.2 at.% at a plateau (50-300 nm depth) plus a presence of oxygen at 9 at.% (see Supplemental Material, figure S6). Even with a base pressure of $6\times10^{-8}$ torr, oxygen incorporation at several at.% level in ScN occurs due to the high reactivity of Sc with oxygen from residual water during deposition [18,32,33,43-46]. Gregoire *et. al.* demonstrated an occupancy of oxygen on the nitrogen site possible from 2 at.% to 6 at.% which correspond at a maximum to $Sc_{0.94}\square_{0.06}N_{0.94}O_{0.06}$ [32]. A higher presence of oxygen in the film leads to the accumulation of oxygen at the grain boundaries and defects [18]. In the present study, the most probable case is an incorporation of oxygen at few at.% level in the $Sc_{1-x}Mg_xN_{1-y}O_y$ (0 < y < 0.06) and $Sc_2O_3$ at the grain boundaries/defects at a few percent (≤ 2 mol.%) (More details can be found in the Supplemental Material, figure S7). This sample with an average of 2.2 at.% of Mg was used as reference in order to calculate the percentage of magnesium in each film from the $Mg^+$ signal intensity detected by TOF-SIMS. It is important to note that the oxygen content does not affect the purpose of the present work, as these oxygen contaminations only marginally affect the thermal conductivity [47]. However, oxygen doping acts as donor doping and leads to a shift $E_F$ towards the conduction band [25,33]. Thus, we do not obtain *p*-type Mg-doped ScN, as in the work of Saha *et al.* [28,35].

From the SRIM simulations, a depth profile of the implanted Mg ions in ScN thin films can be calculated (Figures 3a and 4b). Figure 3a shows how several implantation energies have been used to obtain an approximately constant concentration of magnesium in the ScN film. Figure 3b gives the Mg profile from only one implantation energy (150 keV), with a total dose of Mg equivalent as for the 2.2 RT sample. A maximum of around 3 at.% was expected at ~200 nm from the surface of the sample (figure 3b). The Mg profiles measured with TOF-SIMS are presented in Figure 3c and 3d. Since the thin films have slightly different thicknesses, the film depths were normalized according to film/substrate interface in order to facilitate comparison. The intensities were also normalized to the substrate signal ($Al^+$). The ScN reference sample had no detectable $Mg^+$ signal. The $Mg^+$ signal measured on the flat-profile-implanted samples is slightly lower at the surface of the film, but then almost flat until it drops close to the substrate interface. The intensity of the $Mg^+$ signal is consistent with a higher concentration of magnesium in the film. We note that the sample implanted at high temperature (2.2 HT) and at room temperature (2.2 RT) have the same elemental depth profile features (profile and intensity). A variation of the concentration of Mg is observed along the thickness with a maximum observed at half of the thickness (~200 nm) and, almost symmetrically, a decrease of Mg concentration up to the surface and the film/substrate interface. The profiles throughout the film appear relatively similar for all samples and match the profiles from the SRIM simulations. The small increase of the $Mg^+$ signal appearing at the interface between film and substrate is due to different $Mg^+$ yield in ScN and $Al_2O_3$.

Figure 4 shows HAADF-HRSTEM images of the ScN reference sample, the 2.2 RT sample and 2.2 HT sample where two columnar grains and the grain boundary between them can be observed for both samples. Local Fast Fourier Transforms (FFT) were performed on the different zones marked in the corresponding micrograph. Very similar comments can be addressed for the reference sample (before implantation) (figure 4a) and the high-temperature-implanted sample (2.2 HT) (figure 4c). They both present a high level of ordering and homogeneity inside the grains with sharp spots on local FFT. In other words, these observations did not allow for identification of defects inside the grains which could have formed during the growth process or during the Mg implantation at 600°C. In the case of the room-temperature-implanted sample, the high degree of ordering of the atoms is visible in some parts of the columnar grains and this is confirmed by the sharp spots on the local FFT (see areas #1 and #2 on figure 4b). Blurry and likely defect-rich areas, with a typical size of ten nanometers, are distinguishable as well as a broadening of the spot on their local FFT (area #3). The HRSTEM analysis illustrates the difference between the room-temperature implantation and high-temperature implantation of ScN. By implanting the magnesium at 600°C, the thermal energy during implantation appears to be sufficient to anneal out most of the defects induced by implantation. In contrast, at room temperature, the defects and local misalignment of atoms exists within the grains.

Figure 5a is a closer comparison of the TOF-SIMS analysis of the 2.2 RT and 2.2 SE samples. Both samples had a similar substrate/film interface up to 100-150 nm thickness. At a distance between 150 nm and 300 nm, the 2.2 SE had a higher at.% of Mg (3 at.% locally) than the 2.2 RT . Close to the surface, the 2.2 SE had a lower at.% of Mg than 2.2 RT down to a negligible amount of Mg at the top surface of the film. Figure 5b represents the evolution of total atom displacement (recoil expressed in displacement per atom, dpa) of Sc and N simulated by SRIM in a case of multi-energy and a single-energy implantation aiming for a total dose of $43 \times 10^{15}$ ions/cm$^2$. The two simulated curves show a similar quantity and distribution of displacement per atom (~20-30 dpa) and thus defects induced by the implantation. In terms of composition, the 2.2 RT and 2.2 SE samples differed with different profiles of Mg along the thickness. Nevertheless, in terms of total displacement per atom and total induced defects, the 2.2 RT and 2.2 SE films were essentially identical with an average displacement per atom evaluated at 19 dpa for the sample 2.2 RT and 20 dpa for the 2.2 SE. The different results from TOF-SIMS, HRSTEM, and SRIM simulation are summarized in Table 2.

**Table 2:** The different characteristics of the sample after ion implantations: concentration of dopant and induced defects with their depth profile and average displacement per atom (dpa).

| Sample label | Dopants | | Defects | | |
|---|---|---|---|---|---|
| | Average conc. of Mg (at.%) | Depth profile | Temperature of implantation | Point and/or extended defects | Average dpa along the film |
| 2.2 HT | 2.2 | "flat" | 600 °C | - | - |
| 2.2 SE | 2.2 | Gaussian-like peak | room temperature | yes | 20 |
| 2.2 RT | 2.2 | "flat" | room temperature | yes | 19 |
| 1.1 RT | 1.1 | "flat" | room temperature | yes | 9.5 |
| 0.73 RT | 0.75 | "flat" | room temperature | yes | 6.6 |
| 0.39 RT | 0.35 | "flat" | room temperature | yes | 3.3 |
| REF | 0 | - | - | - | - |

Figure 6 shows the thermal conductivity of the Mg-implanted ScN films. The value of the thermal conductivity of the ScN reference sample is similar to earlier reported values for ScN thin films (10-12 Wm$^{-1}$K$^{-1}$) [25,27,30,47]. The sample implanted at high temperature exhibits a thermal conductivity similar to the value of the reference sample ScN. With a temperature high enough to anneal out the defects, the difference is within the error bars and can thus be considered negligible in this case. The smaller or negligible effect of Mg dopants on thermal conductivity compare to the one observed in previous study with Nb doping can be explain by a lower difference of atomic mass between Sc (44.95 u) and Mg (24.31 u) than Sc and Nb (92.20 u) [27].

For room-temperature implantation, a trend of decreasing thermal conductivity for higher amount of implanted Mg is clear. A large drop between the ScN reference (10.5 Wm$^{-1}$K$^{-1}$) and the sample implanted with 0.37 at.% of Mg (4.2 Wm$^{-1}$K$^{-1}$) can be seen. The other implanted samples at room temperature and using multi-energy implantation have similarly low values of the thermal conductivity as the 0.37 RT sample. A minimum is observed for 0.75 at.% of Mg in ScN with a thermal conductivity of 3.2 Wm$^{-1}$K$^{-1}$. The sample implanted using a single beam-energy also has a similar thermal conductivity, comparable to the lowest observed with a flat Mg concentration-profile.

This large decrease in thermal conductivity (2.5 times lower) for ScN when implanting a small amount of Mg may be explained by the increased level of phonon scattering due to the presence of defects induced by ion implantation. The single-energy implanted sample did not show a substantially different thermal conductivity in comparison to the multi-energy implanted samples. Thus, the Mg concentration-profile along the thickness of the film does not substantially affect the thermal conductivity of the film. In summary, these results indicate that room temperature Mg implantation is preferred if a lower thermal conductivity is desired, to avoid annealing out the defects and retaining the corresponding phonon scattering.

The results of simultaneous measurements of the Seebeck coefficient and the electrical resistivity together with their corresponding power factor are shown in figure 7. Fig. 7 a-c presents the results from the 2.2 at.% Mg sample with three different conditions (room temperature RT, high temperature HT and single-energy SE) plus the as deposited ScN reference sample. In Fig. 7 d-f, the results from the samples with different Mg concentrations are presented. The Seebeck coefficient, the electrical resistivity, and the power factor at certain fixed temperatures as a function of Mg concentration and type of implantation are presented in figure S9 of Supplemental Material.

The ScN reference film is also plotted showing the lowest absolute value of the Seebeck coefficient (-41 µV/K at 775 K). The film implanted at high temperature (5% HT), considered here as "defect-free", exhibited an absolute value of the Seebeck coefficient slightly higher than the ScN reference sample mentioned above (-56 µV/K at 775 K). The trend of increasing the Seebeck coefficient predicted from DFT calculations [33] is corroborated by the results obtained from these experiments. Implantation at room temperature led to samples exhibiting higher absolute values of the Seebeck coefficient up to around -67μV/K (775 K). The Mg concentration profile does not seem to affect the Seebeck coefficient with similar behavior with the temperature for the 2.2 SE. The evolution of the Seebeck coefficient values with the concentration of dopants is low with a maximum absolute values obtained for the sample with 0.75 at.% of Mg (-69μV/K at 775 K). The results from the Seebeck coefficient measurements show at first an effect of the magnesium doping with an increase of the Seebeck values and secondly combining with the creation of defect (point and/or extended defects) another increase of the Seebeck coefficient values.

The lowest electrical resistivity value is observed for the ScN reference sample (~250 μΩ.cm). This sample exhibited almost a constant electrical resistivity values over the whole measured temperature range. The sample implanted at high temperature (2.2 HT) exhibited a temperature dependence of the electrical resistivity ($\rho(T)$) similar to the ScN reference, but with higher values (~750 μΩ.cm). Similar to the Seebeck coefficient, no differences are observed between the multi-energy and the single-energy implanted films. For all the samples implanted at room temperature, a trend of starting with almost constant resistivity values can be observed, but then a decrease after around 450 K. This change of resistivity may be due to recombination of some point defects (such as Frenkel defects). These point defects can recombine at low temperature (a few hundreds of kelvin) and can lead to the creation of extended defects in the materials (line defects such as dislocations or twins). The removal of defects after implantation differs between materials. In the case of silicon, the most studied material for ion implantation, a complete removal of the extended defects can be achieved only at high temperature such as 1100-1300 K [48]. In our case, one can propose that the measurement temperature is insufficient too to anneal the extended defects present before the measurement and/or created by point-defect recombination during the measurements. The temperature dependent resistivity from 70K to room temperature is presented in supplemental material (figure S8) where differences between the reference sample and the 2.2 RT can be observed due to the mobility of charge carrier which is affected by the defects induced during implantation. The resistivity values differ slightly with the concentration of Mg. In the temperature range of measurement, the sample with 0.75 at.% of Mg exhibited the highest values of electrical resistivity and the 2.2 at.% implanted at room temperature the lowest. The increase of the electrical resistivity can be due to a small contribution of Mg insertion into ScN film observed on 2.2 HT sample and an important contribution from the defects created by ion bombardment [49].

The combination of the Seebeck coefficient and the electrical conductivity for ScN reference sample gives the power factor $\sim 0.55 \times 10^{-3}$ W/mK$^2$ at 775 K. The sample implanted at high temperature exhibited a lower power factor than the ScN reference sample as well as the samples implanted with a low amount of magnesium (0.35 to 1.1 at.%). The samples with 2.2 at.% implanted using a single-energy and multi-energies exhibited the highest power factor $0.64 \times 10^{-3}$ W/mK$^2$ (at 775 K).

The lower (absolute) value of the Seebeck coefficient for ScN compared to earlier reported ScN films, is most likely due to the higher amount of oxygen contamination present in the film, especially the presence of oxide at grain boundaries and/or defects [18,19,32]. A presence of oxide at grain boundaries/defects affected the thermoelectric properties with a reduced Seebeck and electrical conductivity resulting to a low power factor [18,27].

Saha *et al* reported on the electrical, carrier concentration and Seebeck coefficient of Sc$_{1-x}$Mg$_x$N films grown by dc-magnetron co-sputtering [28,35]. They reported an increase of electrical resistivity, a decrease of the mobility and room temperature Seebeck values between -50 to -100μV/K when doping with Mg. They also reported a switch from n-type to p-type behavior for film with $x > 0.028$. In the present study, the film contained a higher amount of oxygen and, within the doping range of the study, only n-type behavior was observed. As previously mentioned, the doping by magnesium in ScN shifts $E_F$ towards the valence band, but oxygen doping ScN leads to a shift towards the conduction band [25,33]. The higher oxygen contamination in the present work than in the work of Saha *et al.* thus explains why the n-type behavior is retained also for higher concentration of magnesium in ScN.

Mg doping in ScN with low amount of defects, achieved by high temperature ion implantation, yielded a similar thermal conductivity as the ScN reference and lower power factor due to a higher electrical resistivity. However, implantation of magnesium at room temperature with a constant or Gaussian-like distribution of Mg along the thickness led to samples exhibiting different physical properties. Implantation at room temperature will create point defects and extended defects which play an important role on the conduction of phonons and charge carriers (electrons or holes). Three features can be emphasized here after implantation of Mg: a decrease of the thermal conductivity, an increase of the absolute value of Seebeck coefficient, an increase of the resistivity and a different $\rho(T)$.

## Conclusions

Ion implantation was used to implant Mg in order to induce doping and defects in epitaxial ScN (111) films grown on sapphire substrates using reactive DC magnetron sputtering. $Mg^+$ ions were implanted with different concentration profiles along the thickness of the film. The ion implantations of 0.3 to 2.2 at.% of Mg in ScN did not affect the rock-salt ScN crystal structure nor morphology of the films. A high temperature of implantation tends to anneal the defects, while doping did not alter the thermal conductivity in comparison to a ScN reference ($\approx 10$ $Wm^{-1}K^{-1}$). In contrast, the room-temperature-implanted samples exhibited large reduction in thermal conductivity to values close to 3.2 $Wm^{-1}K^{-1}$ and an increase of the power factor is also observed for the sample with 2.2 at. % Mg compared to the ScN reference samples. Thus, this study showed the importance of ion-induced defects in the material on the thermal conductivity, in that high temperature implantation allows the defects to be annealed out during implantation, while the defects are retained for room-temperature implanted samples, allowing for a drastic reduction in thermal conductivity.

# Acknowledgments

The authors acknowledge the funding from the European Research Council under the European Community's Seventh Framework Programme (FP7=2007–2013) ERC Grant Agreement No. 335383, the Swedish Foundation for Strategic Research (SSF) through the Future Research Leaders 5 program, the Swedish Research Council (VR) under Project No. 2016-03365, the Knut and Alice Wallenberg Foundation through the Wallenberg Academy Fellows program, and the Swedish Government Strategic Research Area in Materials Science on Functional Materials at Linköping University (Faculty Grant SFO-Mat-LiU No. 2009 00971). AS would like to acknowledge DAE-BRNS (37(3)/14/02/2015/BRNS) and IIT Mandi for research facilities.

# Figure 1

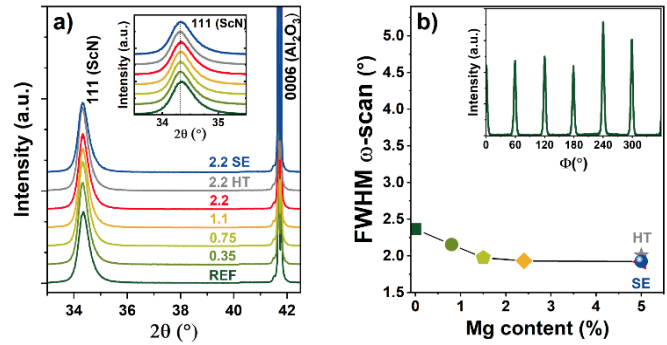

**Figure 1.** a) Offset-separated θ-2θ scans of ScN film grown on *c*-axis-oriented sapphire substrates. The inset graph shows a close up of the ScN (111) peak. The numbers in corresponding colors correspond to the average concentration of Mg in ScN films. b) FWHM values of the rocking curve performed on the 111 reflection. The inset shows ɸ-scan (at ψ =70.5$^0$) of the ScN reference sample grown on sapphire substrate.

# Figure 2

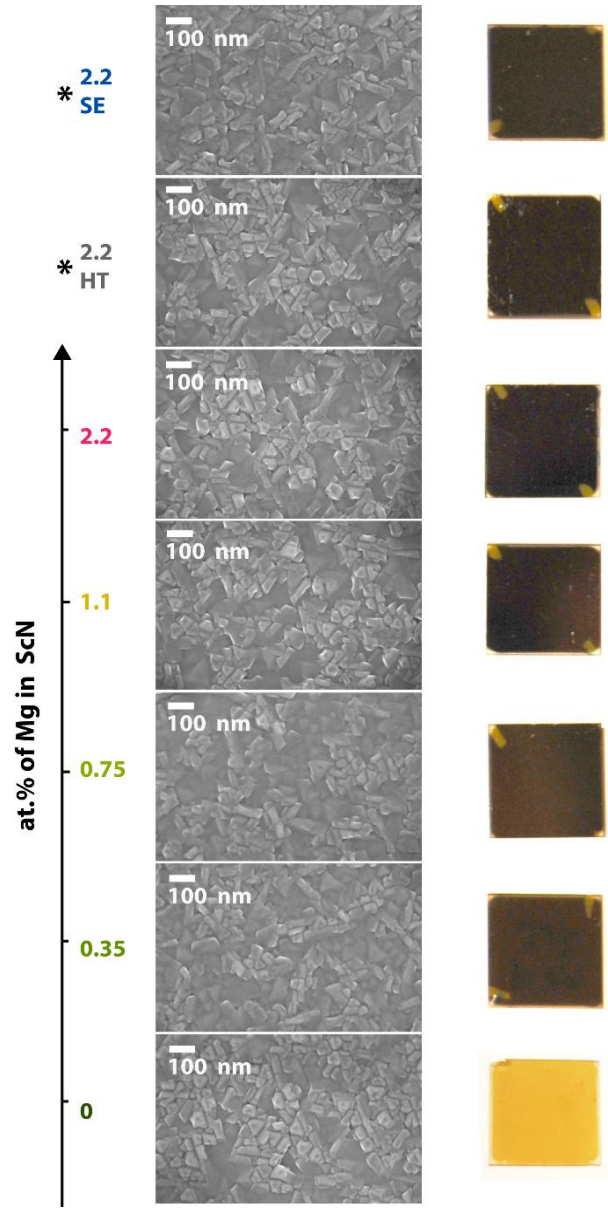

**Figure 2.** The morphology of the Mg-implanted ScN films observed by SEM. To the right, the optical appearance of the films is presented. The numbers to the left indicate the amount of implanted Mg in the ScN films.

**Figure 3**

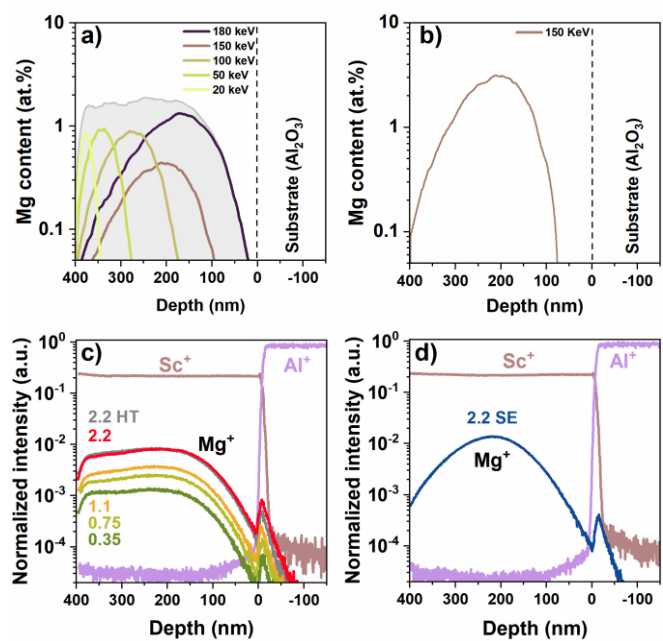

**Figure 3.** a) Simulation results of Mg implantation in ScN using SRIM. Different implantation energies and fluencies were used for a flat Mg profile in the ScN film. The presented graph is based on 2.2 RT sample. b) Simulation of the single-energy implantation (150 keV for the sample 2.2 SE). c, d) TOF-SIMS profiles of selected ions for the different implanted films.

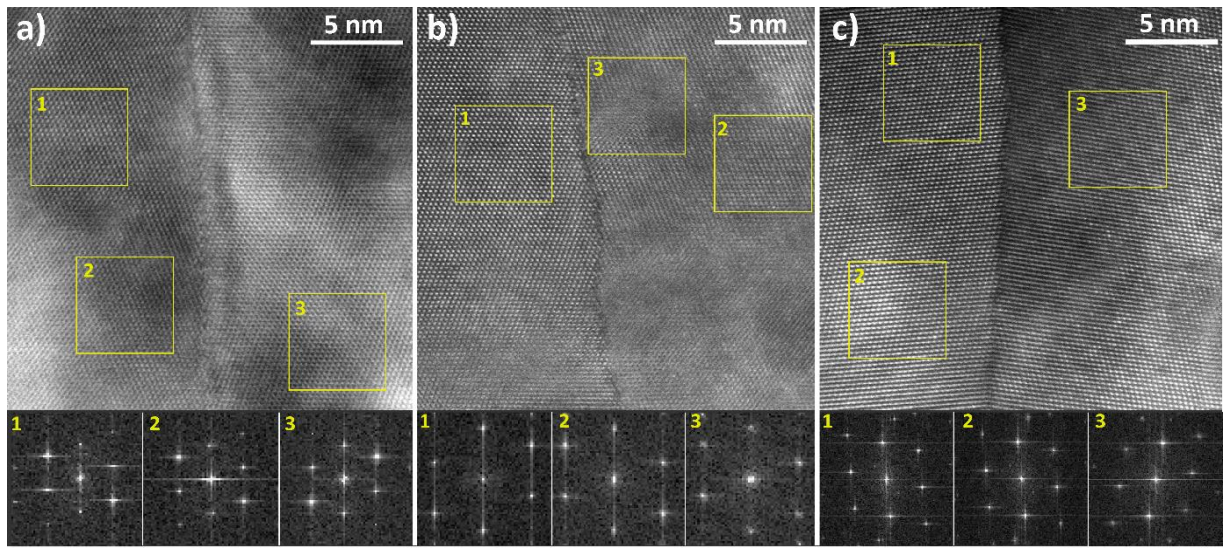

**Figure 4.** HRSTEM micrograph with HAADF detector of the ScN Ref sample (a), the 2.2 RT sample (b) and the 2.2 HT sample (c). Below each image, local FFT of the corresponding zone marked on the image.

**Figure 5**

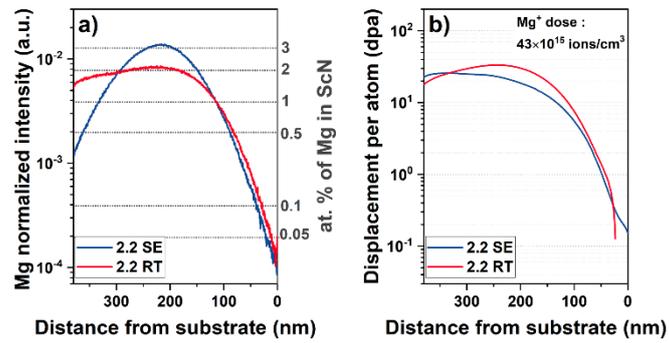

**Figure 5.** Comparison between film implanted with one energy (2.2 SE) or five energies of implantation (2.2 RT): a) TOF-SIMS profile of Mg ions and the estimated at.% of Mg in ScN along the thickness of the film. b) SRIM simulation of the recoil concentration (displacement per atoms) (Sc + N) with a total dose of $Mg^+$ of $43 \times 10^{15}$ ions/cm$^3$.

# Figure 6

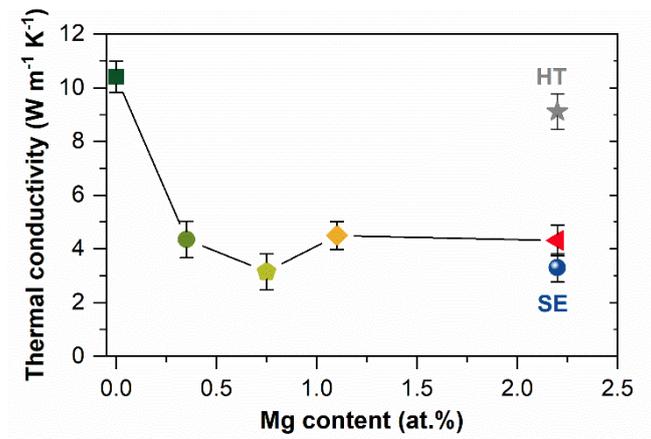

**Figure 6.** The thermal conductivity values of the different alloys obtained by fitting of the modulated thermoreflectance microscopy measurements. Model: 250 nm gold (k= 225 W/mK; D = $0.9\times10^{-4}$ m$^2$/s) / Mg-ScN film on Al$_2$O$_3$ (k = 46 W/mK; D = $1.48\times10^{-5}$ m$^2$/s).

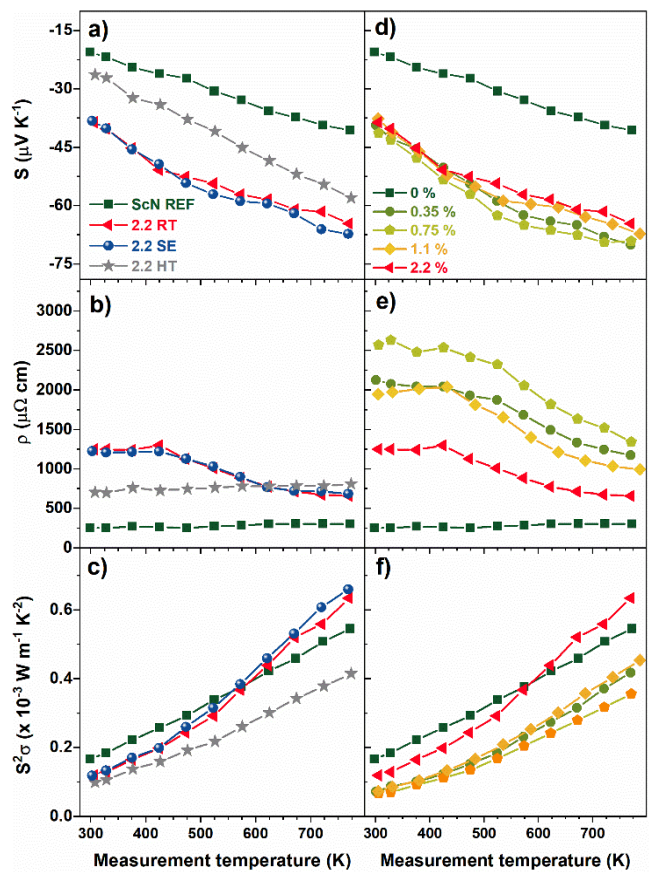

**Figure 7.** The measured Seebeck coefficient (S), the electrical resistivity (ρ) and the power factor ($S^2\sigma$) from room temperature to 770 K of: a, b, c) the reference sample, the 2.2 SE, the 2.2 RT and 2.2 HT; d, e, f) the samples implanted at room temperature with different concentrations of Mg.

# Supplemental material

I. Modulated thermoreflectance setup and analysis

a) Generality and measurement

Thermal conductivity of the films was obtained at room temperature using modulated thermoreflectance microscopy (MTRM). In this setup, a pump beam at 532 nm delivered by a Cobolt MLD laser, intensity modulated by an acousto-optical modulator at a frequency f, is focused on the surface of the sample with an objective lens (N.A. = 0.5). Then, thermal waves were excited in the sample and monitored by the reflectivity surface change recorded around the pump location by another focused laser beam. The specification of the setup is the spatial measurement around the pump beam. We use a 488 nm Oxxius laser to maximize the probe sensitivity to the thermal field on a gold surface. A photodiode and a lock-in amplifier record the AC reflectivity component, in a frequency range between 1 kHz and 1 MHz. The measurement of the reflectivity of the probe on the surface is performed along a x axe from -10 µm to + 10 µm around the pump beam area. The figure S1 represents typical curve of the amplitude and the phase part of the reflectivity signal measured on a gold/substrate and on a gold/film/substrate sample. Finally, the amplitude and phase experimental data were fitted according to a standard Fourier diffusion law to extract the thermal conductivity of the ScN films [38-41].

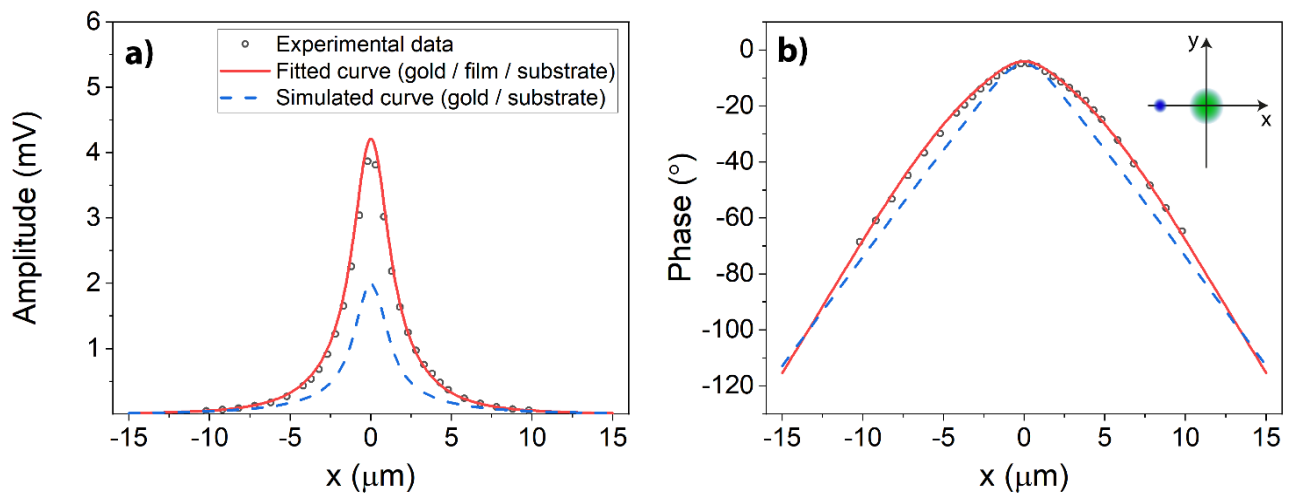

**Figure S1**: a) amplitude and b) phase signals of the reflectivity of the probe beam with the experimental data, the corresponding fitting curve of a gold/film/substrate model and a simulated curve of gold/substrate model for comparison. In inset of b), a schematic view of the measurement along the x axe with the probe (blue) and pump (green) spot.

b) <u>Details on fitting process</u>

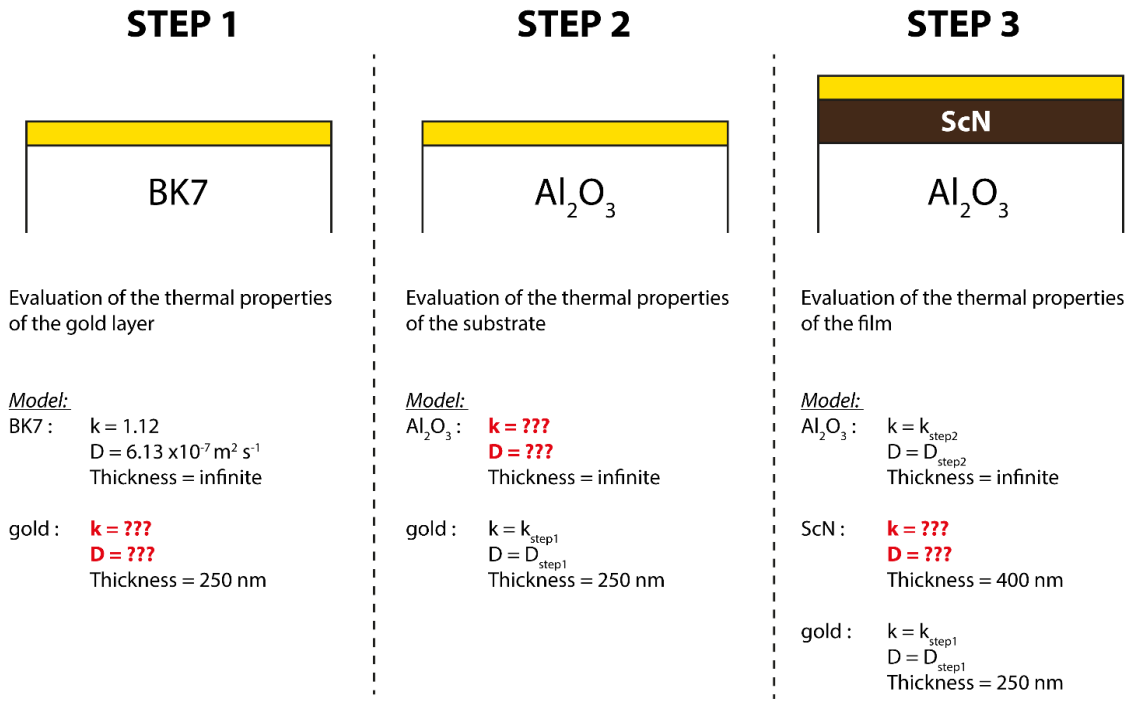

**Figure S2:** Schematic representation of the steps process for the evaluation of the thermal properties of the different layers in the model used for the thermoreflectance measurement. The "???" represents the values fitted in the model to reproduce the experimental values, the other values are considered fixed and known.

In order to remove the difference of the surface aspect such as the reflectivity which may differs in a series of sample, a capping layer composed of was of Au(250nm)/Cr(5nm) was deposited on all samples. the layer of chromium was used as an adhesion layer for gold. Prior deposition, the samples were cleaned with a spray of acetone followed by ethanol to finally be dried with $N_2$ gun. The gold layer is deposited by evaporation on all samples together with the bare substrate and the "BK7" samples. The "BK7" sample is a borofloat® borosilicate glass substrate from Edmund optics ® glass whose thermal conductivity and diffusivity are well known and provided by the supplier. The sample were placed carefully into the thickness homogenized zone of the evaporation chamber holder (2-inch square). Due to the sample configuration the fit is performed with several steps in order to determine the thermal properties of the film. In the heat diffusion equation, the parameters are the thermal conductivity k, the thermal diffusivity D ($D = \frac{\rho . C}{k}$, where ρ is the density of the film, Cp the heat capacity and k the thermal conductivity) and the thickness of each layer. A schematic description of the process is presented on the figure S2.

A first step consists on evaluating the thermal properties (k and D) of the gold layer using a sample gold/BK7 where the thickness of the gold is measured by profilmeter [39]. Once the thermal properties (k and D) of gold capping layer are known, we process to the second step to evaluate the thermal properties of the substrate by fixing in the model the parameters of the gold layer. A best fit allows to evaluate the thermal properties of the substrate which were evaluated to be k= 46 Wm$^{-1}$K$^{-1}$ and a diffusivity of 1.48×10$^{-5}$ m$^2$ s$^{-1}$ [51,52]. Finally, when, the substrates and the capping layer are well known, we continue to the step 3 (described below in detail) with a model including the film in between the capping layer and the substrate.

c) <u>Analysis and fit of experimental data in the case of a sample composed of Gold/ScN/sapphire</u>

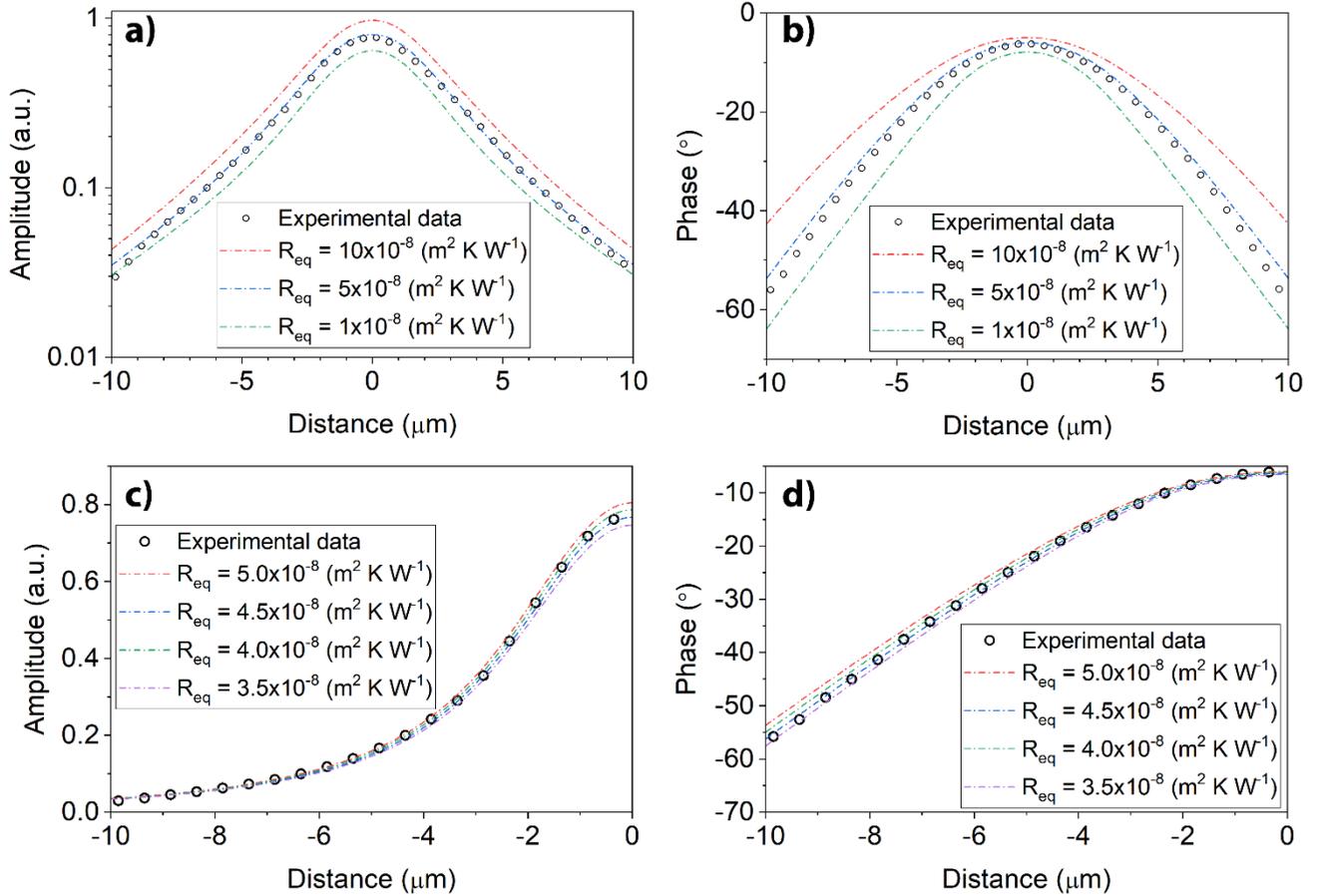

**Figure S3:** spatial amplitude (a and c) and phase (b and d) of the reflected signal of the probe beam (experimental data of the 2.2HT sample) along with the simulated curves varying the total thermal resistant of the film with gold/film/sapphire model

In the case of film with a thermal conductivity lower than the gold (4 -15 Wm$^{-1}$K$^{-1}$ compare to 300 Wm$^{-1}$K$^{-1}$ for the gold), a first approximation is to perform a best fit on experimental data using an equivalent thermal resistant layer whose resistance R$_{eq}$ corresponds to the total film. Once the thermal resistance R$_{eq}$ is determined, the thermal conductivity of the film k$_{film}$ can be calculated by the following equation:

$$R_{eq} = \frac{L_{film}}{k_{film}}$$

Where, L and k represent the thickness and the thermal conductivity of the film respectively. Figure S3 shows the simulated curves for different thermal resistance $R_{eq}$ compared to the experimental data of the 2.2 HT sample presented in the main manuscript. Using the best fit, the experimental data curve can be simulated with a $R_{eq}$ between 4.0 and 4.5 × $10^{-8}$ m² K W⁻¹. At maximum, an error of ± 0.3 $10^{-8}$ m² K W⁻¹ was noticed on the evaluation of the $R_{eq}$ during the "best fit" which correspond to an error of ± 0.6 Wm⁻¹K⁻¹ on the thermal conductivity for this sample. In Figure 6 of the main manuscript, the error observed during the analysis of the sample series is plotted for all the points and varies between ± 0.5 and ± 0.6 Wm⁻¹K⁻¹.

The figure S4 represents the simulated curves with a layer with the thermal conductivity of 9.1 and a diffusivity varying from 1×$10^{-6}$ to 50×$10^{-6}$ m² s⁻¹. the diffusivity of the film has no or a very low impact on the amplitude part of the signal which is more sensitive to the thermal conductivity. Therefore, the diffusivity is deduced from the phase of the signal. The diffusivity of the experimental data can be estimated between 3 and 10×$10^{-6}$ m² s⁻¹ without affecting the estimation of the thermal conductivity. This poor accuracy on the diffusivity of the film does not allow to determine the heat capacity of the film with enough accuracy.

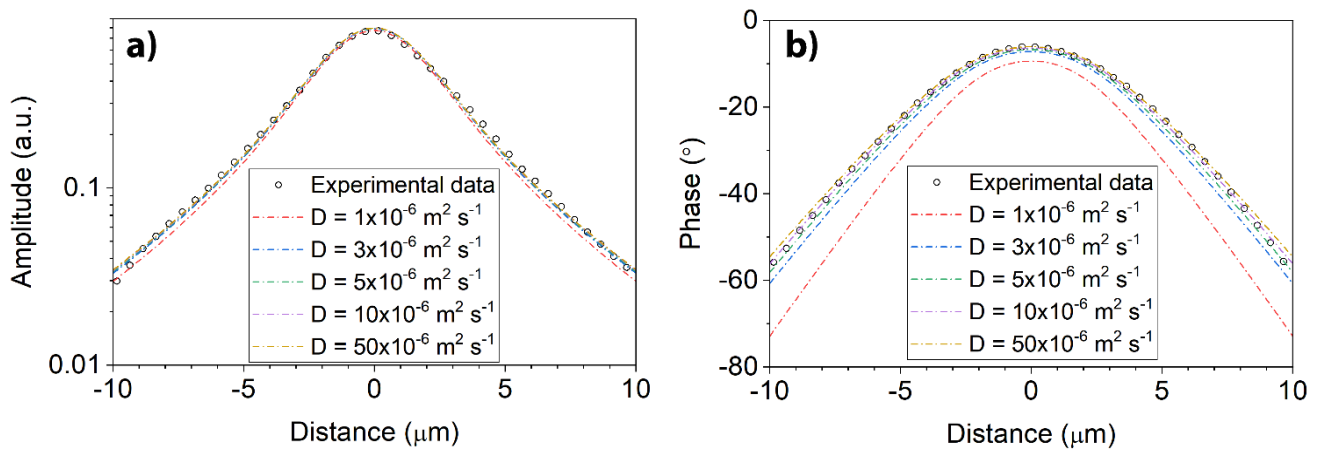

**Figure S4:** spatial amplitude (a) and phase (b) of the measured signal of the sample 2.2HT (experimental data) with simulated curve of a gold/film/ Al₂O₃ model having a film with a k = 9.1 Wm⁻¹K⁻¹ and different diffusivity from 1 to 50 ×$10^{-6}$ m² s⁻¹.

This method allows to evaluate the thermal conductivity of ScN film with an accuracy around ± 0.6 Wm⁻¹K⁻¹ and not the diffusivity (or heat capacity) of the film. In order to determine the diffusivity of the films with high accuracy supplementary measurement on the film without a gold layer on top would be necessary, unfortunately the ScN materials does not absorbed the pump signal used in this setup. Note here that no interface thermal resistant (ITR) layers were used in the model.

d) <u>Interface thermal resistant (ITR) neglected in the system gold/ScN/Al2O3 for evaluating the thermal conductivity</u>

In the previous measurement, the ScN layer was considered as a total thermal resistance and the interface thermal resistance between gold/film and film/Al$_2$O$_3$ were not taken into account. In reality, the thermal resistance is defined as:

$$R_{eq} = ITR_{top} + R_{ScN} + ITR_{bottom}$$

Where ITR$_{top}$ and ITR$_{bottom}$ are the interface gold/film and film/substrate. To evaluate the interface resistance of ScN films, the two ITR needs to be known. A proper way to evaluate those to ITR is to measure thermoreflectance using a series of sample with different film thicknesses and that approximation needs to be assume for a constant ITR regardless the thickness of the film.

A study reported the interface thermal resistance of nitride films where no ITR could not be estimated on the ScN films deposited on MgO due to an extremely low ITR combined to the ScN film masking the ITR [52]. In the same study, the ITR of HfN/MgO, ZrN/MgO and TiN/MgO were evaluated at 5, 3 and 2 $\times 10^{-9}$ m$^2$ K W$^{-1}$. Same approximation can be cone for ITR$_{top}$ where for example gold/Cr ITR values were evaluated to be in the order of $10^{-9}$ m$^2$ K W$^{-1}$ [53]. Approximating the ITR$_{bottom}$ of ScN/Al$_2$O$_3$ in the same order of magnitude as the one observed with the other nitrides, and the ITRtop to the ones observed in a system gold/Cr, the interface thermal resistance would not affect the total resistance of the layer evaluated at 4.5±0.3 $\times 10^{-8}$ m$^2$ K W$^{-1}$.

e) <u>fitting curves of all samples studied</u>

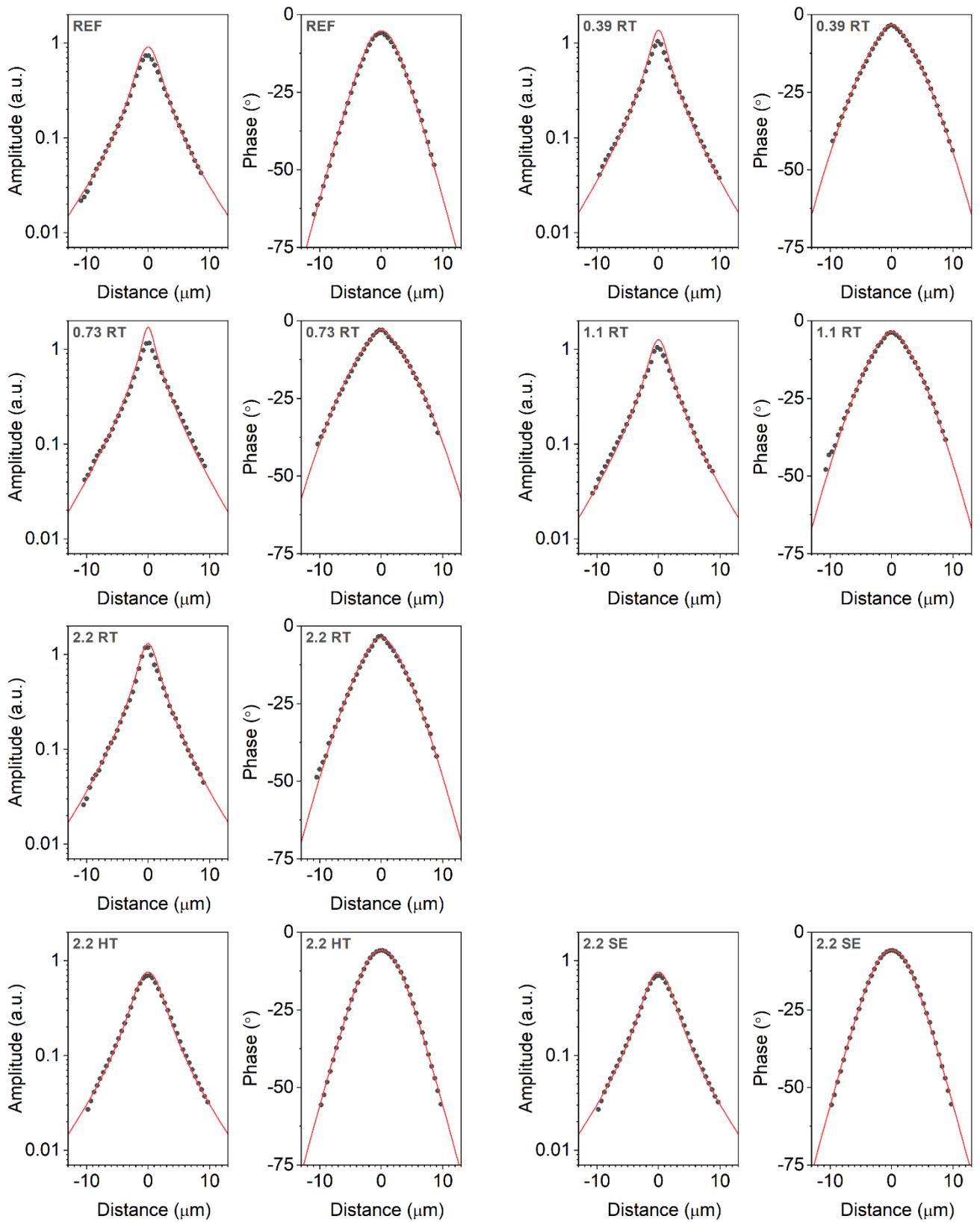

**Figure S5**: a) amplitude and b) phase signals of the reflectivity of the probe beam with the experimental data and corresponding fitting curve of a gold/film/substrate model for all samples. (refer to main article for sample names).

## II. Elemental analysis and description of the composition of the film.

The XPS depth profile analysis of the sample 2.2 RT is presented in Figure S5. Sc, Mg, N and oxygen were detected throughout the thickness of the films with a variation between the 1$^{st}$ and 2$^{nd}$ cycle of sputtering due to surface contamination. We can notice that the scandium concentration is constant (45.0 at.%) and that the nitrogen and oxygen have a reverse evolution and respectively have a concentration at the 10$^{th}$ cycle of 43.7 at.% and 9 at.%. A higher oxygen (lower nitrogen) content is present at the surface. The depth profile of magnesium is consistent with the TOF-SIMS measurement with an increase until a plateau at a 2.15-2.25 at.% of Mg (Mg/(Sc+N+Mg)). This result was used as a reference sample to extrapolate the different atomic concentration in the films using the ToF SIMS Mg intensity detected.

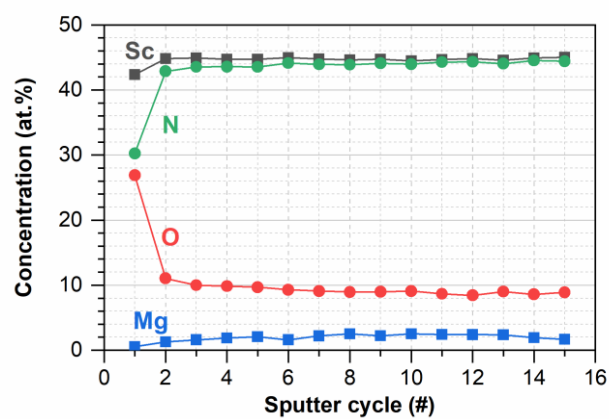

**Figure S6.** Depth profile concentration measured by XPS of the sample 2.2 RT sample.

From XPS measurements, the composition of the film can be determined. The film contains in total 9 % of oxygen which can be incorporated in ScN or like an oxide such as $Sc_2O_3$ at grain boundaries and/or defects. Figure S6 describes the possible compositions of the film from a full incorporation of oxygen in the rock salt structure of ScN at the nitrogen site to a maximum of 3% of $Sc_2O_3$ and 97% of $Sc_{0.89}Mg_{0.05}N_1$. This calculation is base considering a full dissolution of Mg into ScN cell. Dissolved oxygen into ScN in nitrogen site is commonly observed at few at.% level. For example, a 3 at.% level of oxygen in a ScN film leads to $Sc_{0.94}\square_{0.06}N_{0.94}O_{0.06}$. [18,19,25,45] In the present study and following the different papers reporting the growth of ScN, the films are mostly composed of 2 to 3 at.% of oxide present at the grain boundary plus a partial incorporation of oxygen into ScN at few at.% level.

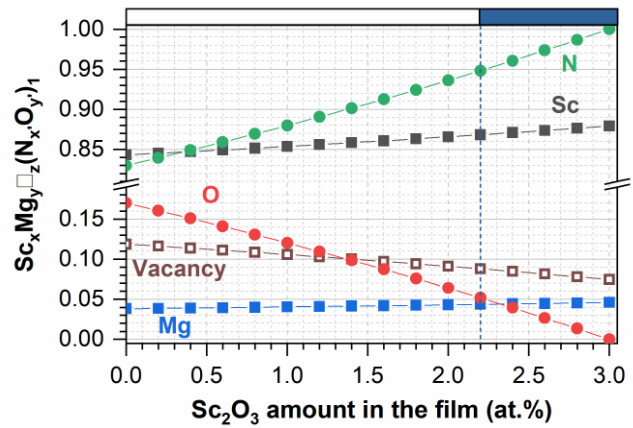

**Figure S7.** Possible composition of $Sc_xMg_y\square_z(N_{x'}O_{y'})_1$ versus the amount of $Sc_2O_3$ present in the film (grain boundaries/defects). The vertical line and the rectangular blue zone at 2.2 to 3 at.% of $Sc_2O_3$ represent the maximum dilution amount of oxygen into ScN reported in the literature (3 at.% of oxygen contamination). Composition deduced from the elemental analysis of the depth profile XPS measurement performed on the sample 2.2 RT.

III. Electrical characterization from low temperature to room temperature

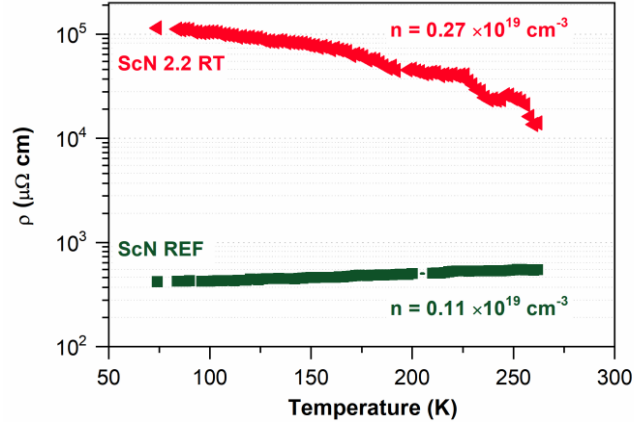

**Figure S8.** Temperature dependence of the electrical resistivity measured from 70 K to room temperature of: a) the ScN REF film and b) and ScN film implanted at room temperature with 2.2 at.% of Mg. the number n correspond to the carrier concentration of the films at room temperature.

The charge carrier concentration measured by Hall-effect measurement on the as deposited ScN sample and the 2.2 RT implanted are $0.11 \times 10^{19}$ cm$^{-3}$ and $0.27 \times 10^{19}$ cm$^{-3}$, respectively. The figure S7 presents the temperature dependence of the electrical resistivity from 70 K to 260 K measured on the ScN reference sample and 2.2 RT implanted. After implantation of 2.2 at.% of Mg in ScN, the electrical resistivity drastically increased from ~550 µΩ.cm to ~10 000 µΩ.cm at 260 K. Temperature dependence of the two films are reverse one has its resistivity increasing slowly when the other one has its resistivity decreasing from 40 K to room temperature. The conduction of electron in a semiconductor or a metal can be scattered by different mechanisms such lattice vibrations, dislocation, impurities, grain boundaries, vacancies etc. The electrical conductivity $\sigma$ defined by:

$$\sigma(T) = n(T).e.\mu_d(T)$$

Where $n(T)$ is carrier concentration, $e$ the electronic charge and $\mu_d(T)$ the drift mobility. Following Mathiessen's rule, the total drift mobility is defined as follow [54]:

$$\frac{1}{\mu_d(T)} = \sum_i \frac{1}{\mu_i(T)}$$

where $\mu_i$ is the drift mobility corresponding to each scattering process involved in the electron conductivity. After implantation of magnesium, a higher carrier concentration $n(T)$ increases and a lower electrical conductivity involves a drift mobility higher and thus an important contribution of defects on the mobility of charge carrier.

## IV. Thermoelectric properties, another view

Figure S8 presents another perspective of the thermoelectric properties with the Seebeck coefficient, electrical resistivity, and power factor. This figure represents the same data as presented in Figure 7, but the evolution of the different characteristics at a constant temperature versus the concentration level of Mg or for different implantation procedure: room temperature/high temperature, multi-energy/single-energy.

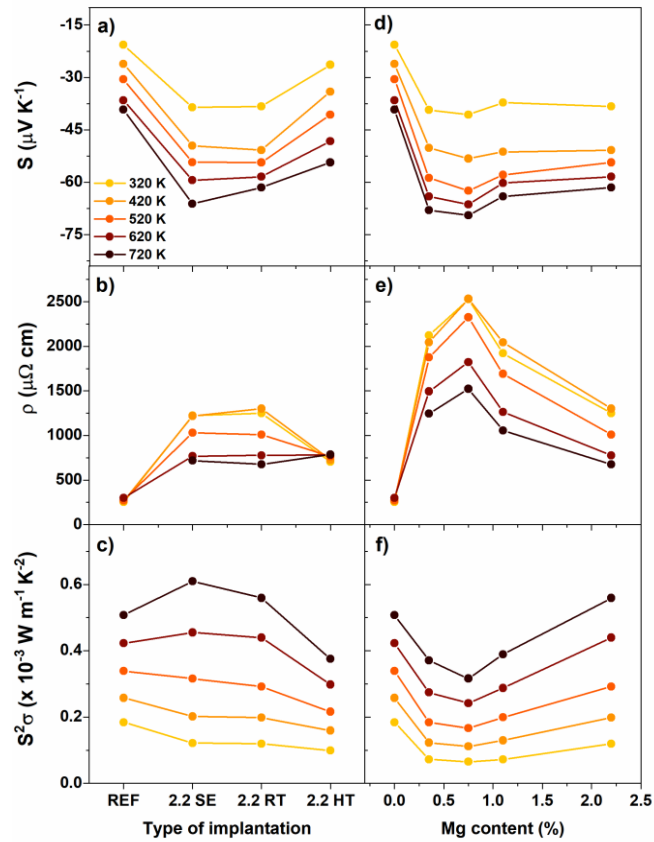

**Figure S9.** The measured Seebeck coefficient (S), the electrical resistivity ($\rho$) and the power factor ($S^2\sigma$) at different temperatures for: a, b, c) the reference sample, the 2.2 SE, the 2.2 RT and 2.2 HT; d, e, f) versus the concentration of Mg implanted into ScN. This figure is a different representation and perspective of the data presented in Figure 7 of main manuscript.

V. High angle annular dark field (HAADF) STEM analysis

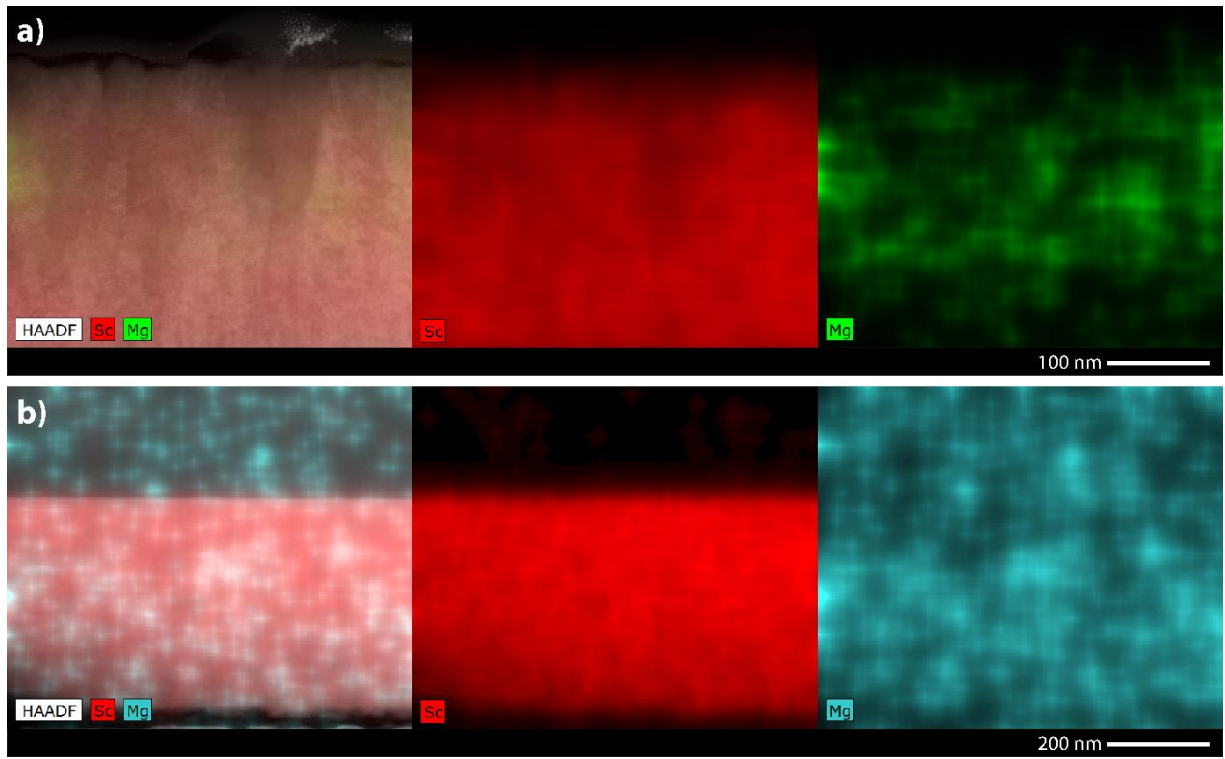

**Figure S10.** HAADF-STEM images of a) the 2.2 RT film and b) 2.2 HT film with the chemical mapping of Sc and Mg. Note here that the signal for collection of magnesium is low and no big contrast and or high concentration zone were observed (grain boundary).